\documentclass{article}

\usepackage{spconf}
\usepackage{amsmath}
\usepackage{graphicx}
\usepackage{makecell}
\usepackage{multirow}
\usepackage{url}

\begin{document}


\title{Resource-efficient Automatic Refinement of Segmentations\\ via Weak Supervision from Light Feedback}

\name{Alix de Langlais$^{1, 2}$,
      Benjamin Billot$^{1}$,
      Théo Aguilar Vidal$^{3, 4}$, 
      Marc-Olivier Gauci$^{2, 3}$,
      Hervé Delingette$^{1}$}

\address{$^{1}$ Inria, Université Côte d'Azur, Epione team, Sophia-Antipolis, France \\
         $^{2}$ Institut de Biologie de Valrose, Inserm U1091, ICARE team, Nice, France \\
         $^{3}$ Institut Universitaire Locomoteur et du Sport, Centre Hospitalier Universitaire, Nice, France \\
         $^{4}$ Université Côte d'Azur, Nice, France}
\maketitle


\begin{abstract}
Delineating anatomical regions is a key task in medical image analysis. Manual segmentation achieves high accuracy but is labor-intensive and prone to variability, thus prompting the development of automated approaches. Recently, a breadth of foundation models has enabled automated segmentations across diverse anatomies and imaging modalities, but these may not always meet the clinical accuracy standards. While segmentation refinement strategies can improve performance, current methods depend on heavy user interactions or require fully supervised segmentations for training. Here, we present SCORE (Segmentation COrrection from Regional Evaluations), a weakly supervised framework that learns to refine mask predictions only using light feedback during training. Specifically, instead of relying on dense training image annotations, SCORE introduces a novel loss that leverages region-wise quality scores and over/under-segmentation error labels. We demonstrate SCORE on humerus CT scans, where it considerably improves initial predictions from TotalSegmentator, and achieves performance on par with existing refinement methods, while greatly reducing their supervision requirements and annotation time. Our code is available at: \url{https://gitlab.inria.fr/adelangl/SCORE}.
\end{abstract}



\section{Introduction}

\begin{figure*}[htb]  
\centering
\includegraphics[width=1\textwidth]{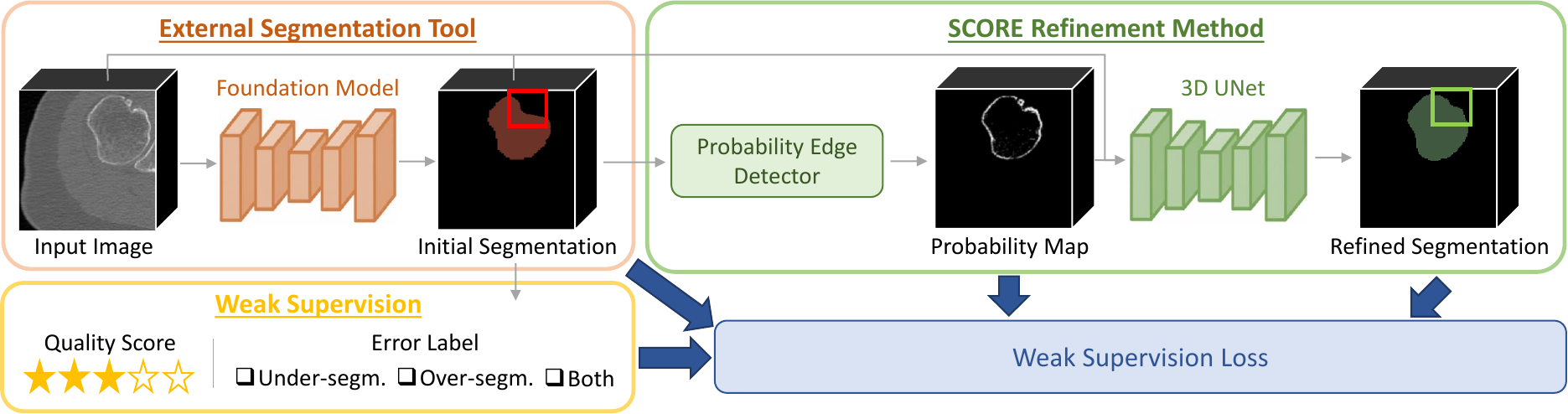}  
\caption{Overview of SCORE, our weakly supervised framework to refine segmentation from an external tool using only light feedback. The network takes as input a 3D image, its initial segmentation, and a probability map for additional edge priors.} \label{fig:pipeline}
\end{figure*}

Modern automated segmentation approaches, best represented by deep-learning networks~\cite{ronneberger_u-net_2015, vaswani_attention_2023}, have been developed to alleviate the labor and variability associated with manual delineation of anatomical regions in medical images. The performance of these networks is intrinsically linked to the scope and variety of their training data, thus prompting the use of large, heterogeneous datasets. This data-centric approach is exemplified by foundation models, such as TotalSegmentator~\cite{wasserthal_totalsegmentator_2023}, which are pre-trained on thousands of multi-source scans to enable robust segmentations across a wide range of input scans. While this strategy yields powerful general-purpose models, their resulting predictions may exhibit local inaccuracies that limit their adoption in clinical applications.

Several methods have been developed to refine automated segmentations. First, semi-automated interactive frameworks propose to place human experts in the correction loop~\cite{isensee_nninteractive_2025}, by allowing them to iteratively correct model predictions. Nevertheless, these systems demand substantial manual effort, thus limiting practical applicability. In contrast, fully automated post-processing techniques aim to remove the need for human intervention. Among these approaches, surface deformation-based methods~\cite{hu_deep_2017} have been proposed to adjust boundaries toward object edges by predicting local displacements. Other fully automated approaches employ secondary refinement networks for correction, either to reclassify voxels within small image patches along the boundary~\cite{tang_look_2021}, or to progressively improve results through a cascaded architecture~\cite{schnider_improved_2023}. Finally, a last class of automatic methods leverages anatomical priors to frame segmentation correction as a denoising problem, where Denoising Autoencoders (DAEs)~\cite{larrazabal_post-dae_2020} or Generative Adversarial Networks (GANs)~\cite{iqbal_generative_2022} are trained to correct segmentation inconsistencies. However, these automated  approaches rely on fully supervised ground-truth (GT) segmentations for training, which are laborious to obtain.

In this paper, we present SCORE (Segmentation COrrection from Regional Evaluations), a weakly supervised framework that learns to refine initial segmentation masks based on light feedback (only used during training). Specifically, we propose to train a refinement network using novel morphology-inspired losses that leverage region-wise quality scores and associated error labels (i.e., under- or over-segmentation). Here, we demonstrate our approach on computed tomography (CT) scans of the humerus, where it substantially improves the accuracy of the TotalSegmentator foundation model~\cite{wasserthal_totalsegmentator_2023}. Overall, SCORE achieves comparable performance to existing semi- and fully automated refinement methods, while requiring neither intensive user interactions nor GT segmentations. These results open perspectives for the adoption of refinement methods in clinical applications.


\section{Methods} 
\label{methodo}

\newlength{\timecolwidth}
\settowidth{\timecolwidth}{25min} 
\begin{table*}[!t]
\centering
\caption{Performance comparison of all refinement methods on each test set in terms of: Dice and HD95 statistics (mean $\pm$ std), training supervision requirements, annotation types, and scan annotation times (both during training and inference). Statistical superiority of SCORE over tested baselines is denoted with $^+$ and inferiority with $^-$ (Wilcoxon signed-rank test, 5\% level).}
\fontsize{9}{10.1}\selectfont
\setlength{\tabcolsep}{2.3pt} 
\begin{tabular}{| l | lll | lll | c | c | >{\centering\arraybackslash}p{\timecolwidth}>{\centering\arraybackslash}p{\timecolwidth}|}  
\hline 
\multirow{2}{*}{\textbf{Methods}} & \multicolumn{3}{c|}{\textbf{Dice}} & \multicolumn{3}{c|}{\textbf{HD95 (mm)}} & \textbf{Training} & \textbf{Annot.} & \multicolumn{2}{c|}{\textbf{Annot. time}} \\
& CHU-Full & CHU-Prx & Wang-Dst & CHU-Full & CHU-Prx & Wang-Dst & \textbf{supervision} & \textbf{type} & Train & Test \\ 
\hline
TotalSegmentator~\cite{wasserthal_totalsegmentator_2023} & 92.4 (0.7)$^+$ & 93.7 (0.8)$^+$ & 91.9 (2.2)$^+$ & 2.1 (0.4)$^+$ & 2.0 (0.2)$^+$ & 5.5 (11.0)$^+$ & - & - & - & - \\ 
 \hline
Manual Correction & 97.9 (0.7)$^-$ & \multicolumn{1}{c}{-} & \multicolumn{1}{c|}{-} & 0.6 (0.1)$^-$ & \multicolumn{1}{c}{-} & \multicolumn{1}{c|}{-} & - & Manual & - & 1h30 \\
nnInteractive~\cite{isensee_nninteractive_2025} & 98.6  (0.2)$^-$ & \multicolumn{1}{c}{-} & \multicolumn{1}{c|}{-} & 0.5 (0.1)$^-$ & \multicolumn{1}{c}{-} & \multicolumn{1}{c|}{-} & - & Semi-autom. & - & 25min \\
Post-DAE~\cite{larrazabal_post-dae_2020} & 96.1 (0.5) & 96.6 (0.5)$^+$ & 94.0 (2.1)$^-$ & 1.5 (0.5) & 1.4 (0.2)$^+$ & 5.4 (10.6) & GT segm. & Automated & 1h30 & 4s\\
Post-DAE+Scan & 97.3 (0.2)$^-$ & 97.6 (0.5)$^-$ & 91.9 (2.0)$^+$ & 0.8 (0.2)$^-$ & 1.1 (0.3)$^-$ & 4.8 (7.6)$^-$ & GT segm. & Automated & 1h30 & 4s \\
SCORE (ours) & 96.2 (0.2) & 97.0 (0.6) & 93.0 (2.0) & 1.4 (0.5) & 1.2 (0.3) & 5.4 (10.9) & Weak & Automated & 4min & 5s \\
 \hline
\end{tabular}
\label{tab:results}
\end{table*}

\noindent\textbf{Problem formulation.} Let $I$ be a 3D image associated with GT segmentation $S$ taking values in $\{1,...,K\}$, where $K$ is the number of regions of interest (ROI). As $S$ is generally not accessible, one can use an automated segmentation model $F$ (e.g., a foundation model) to approximate $S$ with $\tilde{S}=F(I)$. Yet, such models might not always reach accuracy levels required for clinical use. Here, we aim to train a refinement network $\varphi$ to yield segmentations $\hat{S}$ of improved quality. Importantly, instead of training $\varphi$ with previous fully-supervised refinement methods, we introduce SCORE, a new weakly supervised strategy (illustrated in Fig.~\ref{fig:pipeline}) that we describe below. 
\newline

\noindent\textbf{Network inputs and boundary prior.} In order to correct the initial segmentations given by $F$, our network $\varphi$ takes a multi-channel input tensor with three components: (1) the 3D image $I$, to provide anatomical context; (2) the initial segmentation $\tilde{S}$ that we wish to correct; (3) a probabilistic map $P$ of structure contours to serve as a prior to steer the corrections of $\varphi$ towards visible anatomical boundaries. Here, we obtain $P$ by clipping intensities between an Otsu-derived lower bound~\cite{otsu_threshold_1979} and a percentile-based upper bound, and normalizing them in $[0,1]$. Overall, $\hat{S}$ is given by $\hat{S} = \varphi(I,\tilde{S},P)$.
\newline

\noindent\textbf{Weak supervision from quality scores.} SCORE alleviates the needs of previous refinement methods for full supervision by only using light feedback. Specifically, each training pair $(I,\tilde{S})$ is associated with a set of quality scores $\{q_k\}_{k=1}^K$ ranking the accuracy of $\tilde{S}$ for each region $k$ from 0 (unusable) to 5 (excellent). Each quality score $q_k$ is then associated with an error label $l_k\in \{-1,0,1,2\}$ describing the error type for each region: -1 for under-segmentation, 1 for over-segmentation, 2 when a region presents both error types, and 0 for no error (when $q_k$=5). Overall, we hypothesize that these scores are sufficient to train a refinement network in a data-driven way. Moreover, in addition to greatly decreasing annotation time, this light feedback is easily collected using online forms via a simple visualization interface, thus bypassing the need for heavier medical image analysis software.
\newline

\noindent\textbf{Learning strategy.} We train $\varphi$ by designing a morphology-inspired loss function $\mathcal{L}$ that leverages the aforementioned light training feedback. For this purpose, we define a three-fold function that converts the region-wise scores and labels into a voxel-based loss. Specifically, the three components of $\mathcal{L}$ combine two targeted correction losses for over- and under-segmentations ($\mathcal{L}^-$, $\mathcal{L}^+$) together with a regularizing stability loss $\mathcal{L}^\text{stab}$ that penalizes deviations from correctly segmented areas in $\tilde{S}$. We now detail each of these losses.

In this paper, we assume that the foundation model $F$ is relatively accurate, such that its errors lie on the outer parts of $\tilde{S}$. Hence, we first propose a stability loss $\mathcal{L}^\text{stab}$ to encourage the interior parts of $\hat{S}$ and $\tilde{S}$ to be the same. More precisely, if $k$ indexes the ROIs in $\tilde{S}$ and $\hat{S}$, we define these ROI-specific interior areas $\Omega^\text{stab}_k$ as follows: if $l_k \neq 0$, $\Omega^\text{stab}_k$ corresponds to the area covered by $\tilde{S}_k$ eroded by a factor $\eta$; otherwise ($l_k=0$), since $\tilde{S}_k$ is excellent, we extend the stability area $\Omega^\text{stab}_k$ to the whole area occupied by $\tilde{S}_k$. As such, we define $\mathcal{L}^\text{stab}$ as a binary cross-entropy between $\hat{S}_k$ and $\tilde{S}_k$ inside $\Omega^\text{stab}_k$:
\begin{equation}
\mathcal{L}^\text{stab} = \frac{1}{K} \sum_k \frac{1}{\Omega_k^\text{stab}} \sum_{v\in \Omega_k^\text{stab}} -\tilde{S}_{k,v} \log(\hat{S}_{k,v}), 
\label{eq:loss_stab}
\end{equation}
where $v$ indexes the coordinates in 3D tensors like $\hat{S}_k$ and $\tilde{S}_k$.

Next, we correct regions with under-segmentations (i.e., $l_k\in \{-1,2\}$) by using an expansive loss $\mathcal{L}^+$ that encourages $\varphi$ to add voxels to $\tilde{S}_k$. Unlike $\mathcal{L}^\text{stab}$, $\mathcal{L}^+$ is applied to the outer part of $\tilde{S}_k$ (named $\Omega^\text{corr}_k$), defined as the band obtained by (1)~dilating and eroding $\tilde{S}_k$ by a factor $\eta$, (2)~subtracting the eroded mask from the dilated one. Importantly, abusive expansions of $\tilde{S}_k$ are prevented with a dual-mechanism. First, we ensure that new voxels are only added in anatomically plausible locations by weighting $\mathcal{L}^+$ with the boundary probability map $P$. 
Second, we use correction weights $w_k\!=\!(5\!-\!q_k)/5 \in [0,1]$ to adapt the magnitude of $\mathcal{L}^+$ to the quality scores, where $w_k\!=1\!$ inflicts full penalty, and $w_k\! =\! 0$ means no correction. If $|\;|$ is the cardinality operator, we have:
\begin{equation}
\mathcal{L}^+ = \frac{1}{K} \sum_k\frac{1}{|\Omega_k^\text{corr}|} \sum_{v\in \Omega_k^\text{corr}} -w_kP_v \log(\hat{S}_{k,v}).
\label{eq:loss_add_new}
\end{equation}

Finally, we tackle over-segmentation errors (i.e., $l_k\in \{1,2\}$) using a subtractive loss $\mathcal{L}^-$ that encourages the removal of boundary voxels. As before, $\mathcal{L}^-$ operates in $\Omega^\text{corr}_k$, and is also weighted by $w_k$. However, we modify the boundary probability weighting by now considering $(1-P_v)$ to preferentially remove voxels away from the contours:
\begin{equation}
\mathcal{L}^- = \frac{1}{K} \sum_k \frac{1}{|\Omega_k^\text{corr}|} \! \sum_{v\in \Omega_k^\text{corr}} -w_k (1-P_v) \log(1-\hat{S}_{k,v}).
\label{eq:loss_rem_new}
\end{equation}

Overall, if $\lambda^\text{stab}$, $\lambda^+$, and $\lambda^-$ denote tunable coefficients, $\varphi$ is trained using a total loss $\mathcal{L}(\tilde{S}, \hat{S}, \{q_k\}, \{l_k\}, P)$ given by:
\begin{equation}
\mathcal{L} = \lambda^\text{stab}\mathcal{L}^\text{stab} + \lambda^+\mathcal{L}^+ + \lambda^-\mathcal{L}^-.
\label{eq:total_loss_new}
\end{equation}

\noindent\textbf{Data augmentation.} To enhance the robustness of SCORE, we develop a comprehensive data augmentation pipeline that simulates variability in input images $I$ and initial segmentations $\tilde{S}$. First, we apply standard augmentations to the input 3D image $I$ in the form of intensity perturbations (Gaussian blurring, Gaussian noise, contrast corruption). Then, we spatially deform the training pairs ($I$, $\tilde{S}$) using random affine transforms and flips in the left-right direction. Finally, we augment ($\tilde{S}_k$, $q_k$) pairs using score-guided morphological augmentation to increase robustness against different ``strengths'' of segmentation errors. For this purpose, we degrade under- and over-segmentations by respectively applying random erosions and dilations. Importantly, these operations are applied non-uniformly using spatially-varying 3D erosion/dilation kernels, whose radii at each location $v$ are determined by a smooth random field. Quality scores $q_k$ are then updated based on the volumetric changes in $\tilde{S_k}$. 
\newline  

\noindent\textbf{Implementation details.} $\varphi$ is implemented as a 3D UNet~\cite{ronneberger_u-net_2015}, and trained with the Adam optimizer~\cite{kingma_adam_2017} (1e-4 learning rate, batch size of 1). The final model is selected based on a validation set, which includes scans of diverse fields-of-view (FoVs) to represent varying clinical acquisition setups. Finally, all hyperparameters ($\eta=2$, $\lambda^\text{stab}=5$, $\lambda^+=1$, and $\lambda^-=1$) are fine-tuned on the validation set.


\section{Experiments and Results}

\begin{figure*}[t]  
\centering
\includegraphics[width=0.98\textwidth]{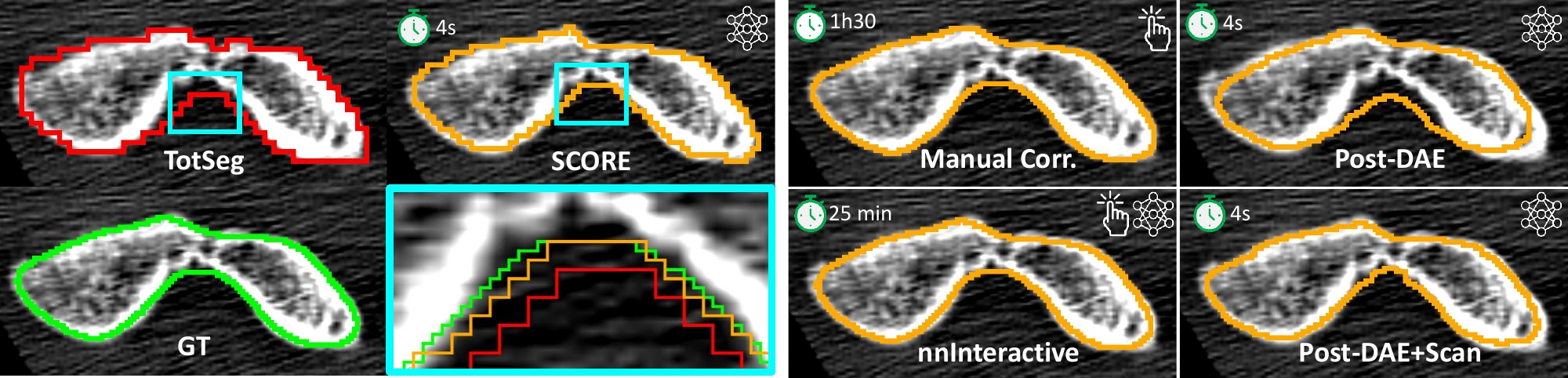}  
\caption{Example segmentation for all methods on a CHU-Full subject, with associated correction time and interaction type.}
\label{fig:resultats}
\end{figure*}

\noindent\textbf{Datasets.} In this paper, we apply SCORE to refine CT segmentations of the humerus given by TotalSegmentator~\cite{wasserthal_totalsegmentator_2023}. SCORE is trained on a private dataset of 31 humerus scans from CHU Nice (FoVs of 512×512×343 to 768×768×1417 voxels; resolutions of 0.23×0.23×0.3 to 0.61×0.61×1.5~mm$^3$). We then use three test sets. The first two are private datasets from CHU Nice: \textbf{CHU-Full} with 5 complete humeral scans of similar FoVs and resolutions as before; \textbf{CHU-Prx} with 35 proximal (i.e., shoulder end~\cite{mostafa_anatomy_2025}) scans with the same resolutions but smaller FoVs (512×512×[158-662] voxels). The third test set, \textbf{Wang-Dst}, is a public dataset of 85 distal (i.e., elbow end) scans with FoVs of 512×512×[188-497] voxels, and resolutions of 0.24×0.24×0.5 to 0.78×0.78×1~mm$^3$~\cite{wang_comprehensive_2024}. 
\newline

\noindent\textbf{Baselines.} We first compare SCORE against \textbf{manual correction} and \textbf{nnInteractive}~\cite{isensee_nninteractive_2025}, the state-of-the-art in semi-automated correction methods. Then, we assess \textbf{Post-DAE}~\cite{larrazabal_post-dae_2020}, the state-of-the-art method in fully automated refinement. In addition to its original configuration (which only takes $\tilde{S}$ as input), we extend Post-DAE into \textbf{Post-DAE+Scan}, a variant that also uses $I$ as input for improved context.
\newline

\noindent\textbf{Results.} As shown in Table~\ref{tab:results}, SCORE delivers statistically significant improvements over the initial TotalSegmentator predictions across all test sets. On CHU-Full, our method increases the mean Dice from 92.4\% to 96.2\% and reduces the HD95 from 2.1 to 1.4 mm. This performance generalizes to the other datasets, with significant Dice improvements of 3.3\% and 1.1\% on CHU-Prx and Wang-Dst, respectively. 

Interestingly, SCORE performs on par with the fully supervised methods Post-DAE and Post-DAE+Scan. Compared to Post-DAE, SCORE shows no statistical difference on CHU-Full, is superior on CHU-Prx, and yields the same HD95 but inferior Dice on Wang-Dst. While Post-DAE+Scan performs better on both CHU test sets, SCORE significantly outperforms it on Wang-Dst in Dice. While these methods perform very similarly in terms of accuracy, we note that SCORE only needs weak labels for training, here enabling us to reduce the annotation time by approximately 95\%.

Finally, manual correction and the semi-automated nnInteractive method yield the highest accuracies, with respective Dice improvements of 1.7\% and 2.4\% over our method on CHU-Full. However, these slight performance gains come at prohibitive annotation costs (25 and 90 minutes per scan, respectively), thus preventing deployment at scale, such as on CHU-Prx (n=35) and Wang-Dst (n=85). In contrast, SCORE only requires about 4 minutes to gather the necessary feedback for training, positioning it as a practical solution to balance accuracy with annotation efficiency.
\newline

\begin{table}[t]
\centering
\caption{Dice statistics (mean, std) for the ablation study.}
\fontsize{9}{10.1}\selectfont
\setlength{\tabcolsep}{8pt}
\begin{tabular}{| l | c | c | c |} 
\hline
\textbf{Methods} & \textbf{CHU-Full} & \textbf{CHU-Prx} & \textbf{Wang-Dst} \\
\hline
SCORE (ours) & \textbf{96.2 (0.2)} & \textbf{97.0 (0.6)} & \textbf{93.0 (2.0)} \\
\hline
$l_k \in \{-1,0,1\}$ & 95.1 (0.4) & 96.0 (0.9) & 92.4 (2.0) \\
$\ominus$ $\mathcal{L}^\text{stab}$ & 54.6 (3.9) & 36.1 (10.4) & 65.6 (7.9) \\
$\ominus$ $P$ & 93.9 (0.6) & 95.4 (0.9) & 91.8 (2.1) \\
$\ominus$ morph. augm. & 95.9 (0.2) & 96.4 (0.8) & 92.3 (2.1) \\
\hline
\end{tabular}
\label{tab:abstudy}
\end{table}

\noindent\textbf{Ablation study.} We now assess design choices in SCORE (Table~\ref{tab:abstudy}). First, ablating multi-class labels (i.e., $l_k$ cannot be 2) shows the benefits of accounting for several errors in each $\tilde{S}_k$. By ensuring consistency in interior parts, $\mathcal{L}^\text{stab}$ greatly contributes to accurate results. Then, removing boundary priors (i.e., $P$) leads to a small decrease in accuracy. Finally, our proposed non-homogeneous morphological augmentation of ($\tilde{S}_k$, $q_k$) pairs slightly increases the robustness of SCORE.


\section{Conclusion} 
In this paper, we present SCORE, a weakly supervised framework for refining errors in initial segmentations from foundation models. Our core contribution is a morphology-inspired three-fold loss that translates region-specific quality scores and error-type labels into voxel-wise corrections. Our results show that SCORE improves initial masks from TotalSegmentator to a performance level comparable to fully supervised methods, while reducing training annotation requirements by 95\%. Future work will explore SCORE generalizability across other anatomies and foundation models. Overall, SCORE offers a path toward making automated segmentation refinement practical and scalable for routine clinical use.
\newline


\noindent\textbf{Acknowledgments.} This work is funded by an Inserm–Inria PhD fellowship.

\noindent\textbf{Compliance with ethical standards.} This work is a numerical simulation study that did not require ethical approval. 

{\fontsize{9}{9}\selectfont
\bibliographystyle{IEEEbib}
\bibliography{references}}

\end{document}